\documentclass[9pt,twocolumn,twoside]{osajnl}
\usepackage[per=slash, loctolang={UK:english}, range-units=single, range-phrase=--, separate-uncertainty]{siunitx}
\usepackage{comment}
\DeclareSIUnit{\rpm}{rpm}
\DeclareSIUnit{\fps}{fps}
\DeclareSIUnit{\pixel}{pixel}
\journal{ao} 

\setboolean{shortarticle}{false} 

\title{A step-by-step guide to reduce spatial coherence of laser light using a rotating ground glass diffuser}

\author[1]{Tim Stangner}
\author[1]{Hanqing Zhang}
\author[1]{Tobias Dahlberg}
\author[1]{Krister Wiklund}
\author[1,*]{Magnus Andersson}

\affil[1]{Department of Physics, Umeå University, 901 87 Umeå, Sweden}

\affil[*]{Corresponding author: magnus.andersson@umu.se}

\ociscodes{(120.4820) Optical systems; (140.3460) Lasers; (230.1980) Diffusers; (110.0110) Imaging systems; (110.4980)  Partial coherence in imaging;}

\begin{abstract}
Wide field-of-view imaging of fast processes in a microscope requires high light intensities motivating the use of lasers as light sources. However, due to their long spatial coherence length lasers are inappropriate for such applications as they produce coherent noise and parasitic reflections, such as speckle, degrading image quality. Therefore, we provide a step-by-step guide for constructing a speckle-free and high contrast laser illumination setup using a rotating ground glass diffuser driven by a stepper motor. The setup is easy to build, cheap and allows a significant light throughput of \SI{48}{\%}, which is \SI{40}{\%} higher in comparison to a single lens collector commonly used in reported setups. This is achieved by using only one objective to collect the scattered light from the ground glass diffuser. We validate the stability and performance of our setup in terms of image quality, motor-induced vibrations and light throughput. To highlight the latter, we record Brownian motion of micro-particles using a 100x oil immersion objective and a high-speed camera operating at \SI{2 000}{\hertz} with a laser output power of only \SI{22}{\milli\watt}. Moreover, by reducing the objective magnification to 50x sampling rates up to \SI{10 000}{\hertz} are realized. To help readers with basic or advanced optics knowledge realizing this setup we provide; a full component list, 3D-printing CAD files, setup protocol, and the code for running the stepper motor.
\end{abstract}

\setboolean{displaycopyright}{true}

\begin{document}

\maketitle

\section{Introduction}
\label{sec:Introduction}
Visualizing fast processes such as particle, cell or protein diffusion using high-speed cameras, necessitates light sources with high intensity output. Lasers fulfill this criterion as they produce coherent beams with high light intensities. Though in terms of sample illumination, their long spatial coherence length is problematic, since coherent noise such as subjective speckles introduces artifacts in microscopic images. In this context, subjective speckles refer to a random granular pattern originating from interference between highly coherent light and its back reflections from rough surfaces, such as glass \cite{DaintyJ.C.1975,Goodman1976,Goodman2005}. In presences of subjective speckle pattern, image quality is poor making it impossible to reveal detailed information about the system under study \cite{Gratton2006}.

\begin{figure*}[t]
\centering
\includegraphics[width=\textwidth]{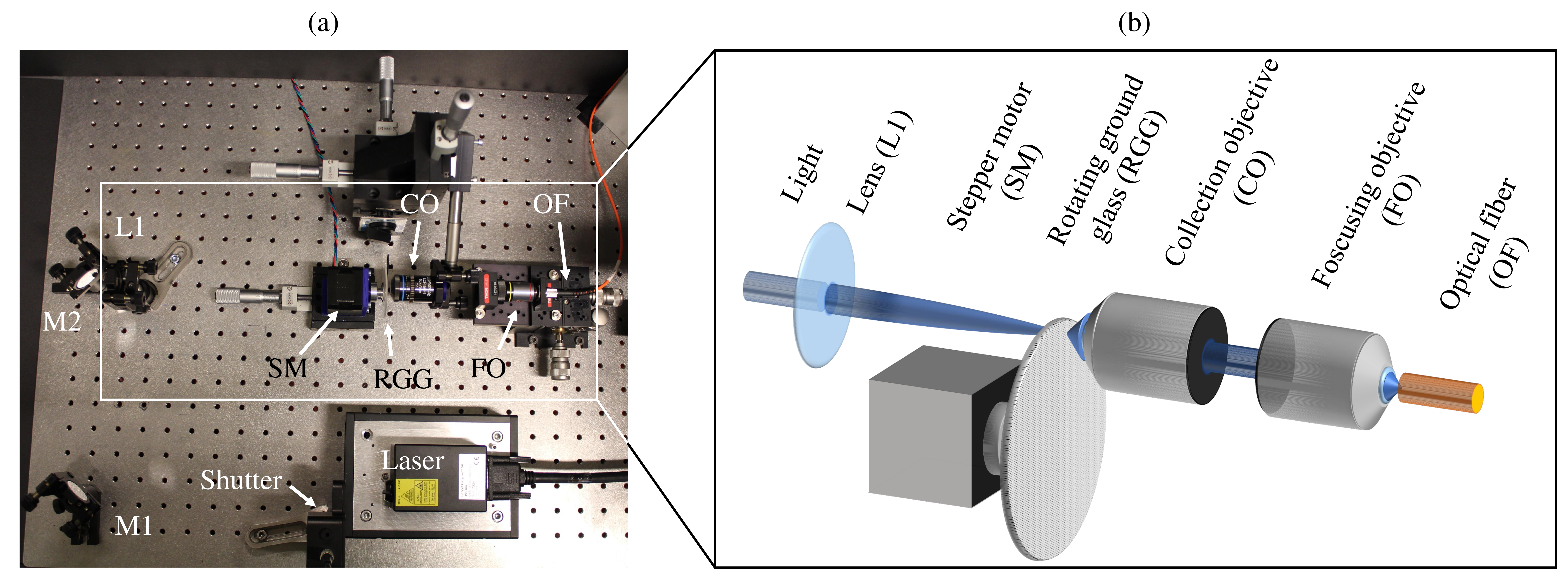}
\caption{Photograph (a) and schematic (b) of the speckle-free laser illumination setup. The laser beam is centered on a $f=\SI{100}{\milli\meter}$ lens (L1) by two mirrors (M1, M2), focusing the light on the rotating ground glass diffuser (RGG). Rotation of the RGG is accomplished by using a stepper motor (SM). The scattered light from RGG is collected with the collection objective (CO), producing a parallel beam output at its back aperture. Subsequently, the light is focused into an optical fiber (OF) using a focusing objective (FO).}
\label{fig:Setup}
\end{figure*}
Besides using high intensity light sources with low spatial coherence length such as superluminescent diodes (SLD) \cite{alphonse1989super,Goldberg1994}, laser driven light sources \cite{smith2011laser} or random lasers \cite{Redding2012}, speckles can effectively be removed by reducing the spatial coherence of laser light sources. To realize the latter, several methods have been suggested including, induction of vibrations in a section of an optical fiber \cite{Ellis1979}, scrambling the light by focusing the laser beam into a multi-mode optical fiber \cite{Davenport1992} or guiding the laser beam on a holographic diffuser \cite{Bains1993}. However, one of the most common approach to remove coherent noise is positioning a rotating or vibrating ground glass diffuser in the beam path, producing a partially coherent light source \cite{Asakura1970,Estes1971,Kumar2010,Zhai2005,Dubois2004,Fougeres1994,Neil2000,Scarcelli2004,Wang2006,Shapiro2008,Ferri2005,Zhang2007,Funatsu1995}. To collect the scattered light from a diffuser single lenses with different focal lengths \cite{Dubois2004,Fougeres1994,Neil2000,Scarcelli2004,Wang2006} or beam splitter cubes \cite{Shapiro2008,Ferri2005,Zhang2007} are used in literature, offering only modest light collection. Additionally, by using a single lens as collector, the width of the collimated beam is in the order of the lens diameter, typically \SI{25.4}{\milli\meter}, and needs to be reduced afterwards by adding additional lenses and therefore increasing alignment effort significantly.

Even-though some of these approaches to reduce the spatial coherence of laser light are published 40 years ago, replicating such setups remains challenging since experimental details are rarely published. Especially, details about motor choice and control, how to mount the ground glass diffuser on the motor, motor rotation speed and how to ensure a high light throughput, are crucial to assemble a robust setup but also to keep construction time and cost low.

To overcome issues with motor choice and control, commercial solutions are available such as motorized rotation stages. However, these stages are made to achieve high angular precision, usually only offering rotation speeds in the range of \SIrange{0.1}{3}{\hertz} which is to slow for effective speckle averaging at short camera shutter times. In addition, these precision motorized rotation stages and their respective drivers are expensive with prices up to several thousands of USD.

Therefore, we provide a simple guide to build a cheap and robust speckle-free laser illumination setup providing a high light throughput to realize camera sampling rates up to \SI{10 000}{\hertz} using only laser powers in the low \SI{}{\milli\watt}-range. The centerpiece of the setup is a rotating ground glass diffuser mounted on a stepper motor and one objective to collect scattered light behind the diffuser. The total cost for stepper motor, its electrical controller unit and objective is below 500 USD. We document each construction and alignment step and characterize our setup in terms of image quality, light throughput and motor-induced vibrations. To highlight a light throughput of \SI{48}{\%}, we investigate the Brownian motion of micro-particles using a high-speed camera. Additionally, we provide; a full component list, CAD-files for constructing mounts and holders and a code to control stepper motors, allowing readers to upgrade the design to fit their experimental requirements. 
\section{Materials and Methods}
\label{sec:MAM}
\subsection{Building a speckle-free laser illumination setup using a stepper motor and a rotating ground glass diffuser}
\label{sec:Setup}
The speckle-free laser illumination setup (Fig. \ref{fig:Setup} a,b) is realized by focusing a laser beam on a rotating ground glass diffuser and coupling the scattered light into an optical fiber. As light source, we use a single frequency continuous wave (CW) diode pumped laser (Cobolt 04-01 Series, Calypso\texttrademark 50, $\lambda= \SI{491}{\nano\meter}$, max. output power $\SI{45}{\milli\watt}$, Cobolt AB, Solna, Sweden), possessing good optical properties such as low intensity noise, high pointing stability and a coherence length of $\approx\SI{300}{\meter}$. In general, however, any laser can be used as light source since the speckle pattern depends only on the angle, polarization, and wavelength of the illuminating laser beam \cite{Trisnadi2002}. In the first alignment step, we couple the laser light, using an focusing objective (FO), into the optical fiber (OF) that is mounted on a \textit{xyz}-stage of a fiber launch system. To illuminate the back aperture of the objective (FO), which focus the incoming light into the fiber, we walk the beam using two mirrors (M1, M2). We optimize the fiber alignment by monitoring the output power via a power meter (351 Hand-Held Power Meter, UDT Instruments (Gamma Scientific)) to ensure maximum coupling efficiency. Furthermore to achieve optimal coupling conditions, the numerical aperture (N.A.) of the fiber has to be equal or bigger than the N.A. of the focusing objective. 

Second, we mount the collection objective (CO), the objective that collects scattered light from the diffuser, on a kinematic mirror mount which is hold by a \textit{xyz}-translation stage. Afterwards, this objective is placed into the beam path, with its front aperture facing the incoming light (details see section \ref{sec:Appendix}\ref{subsec:InstallationDetails}, Fig. \ref{fig:AppendixRGG}). By tilting the objective using the kinematic mirror mount and moving it with the \textit{xyz}-translation stage, we ensure that the back reflection from the front aperture goes back to the light source. However, the back reflection should not hit the laser cavity to avoid strong intensity fluctuations. Furthermore, the correct position of the CO guarantees that the output beam is centered on the back aperture of the focusing objective (Fig. \ref{fig:Setup} b). Optimal alignment is achieved if the fiber output reaches its maximal value on the power meter.

In the third step, we introduce the stepper motor (SM) with the mounted ground glass diffuser (RGG) on its shaft in front of the collection objective. Again, back reflections from the RGG surface are centered on the laser. In this context, it must be highlighted, that the grit-polished side of the RGG must face away from the source. Subsequently, we set the distance between ground glass diffuser and collection objective to its working distance by using the translation stage of the motor (Table \ref{tab:ComparisonObjective}). This step is eminent to ensure parallel beam output on the back aperture of the collection objective.

Eventually, we place a $\SI{100}{\milli\meter}$ lens (L1) between mirror M2 and RGG to focus the light on the latter. This is necessary if the scattered light is collected with an objective or a single lens. We will discuss this issue in section \ref{sec:ValidationExp}\ref{subsec:OutputIntensity}. After placing each optical component in the beam path, we fine tune the optical alignment by slightly changing the positions of the collection objective (in \textit{xyz} and tilt) and of the fiber holder (in \textit{xyz}) to place the focus of the focusing objective into the fiber entrance. We repeat this procedure until the maximum output power is obtained on the power meter. 
\subsection{Controlling the stepper motor using the Arduino Uno micro-controller and the motor driver module A4988}
\label{sec:Motor}
\begin{figure}[hbtp]
\centering
\includegraphics[width=\linewidth]{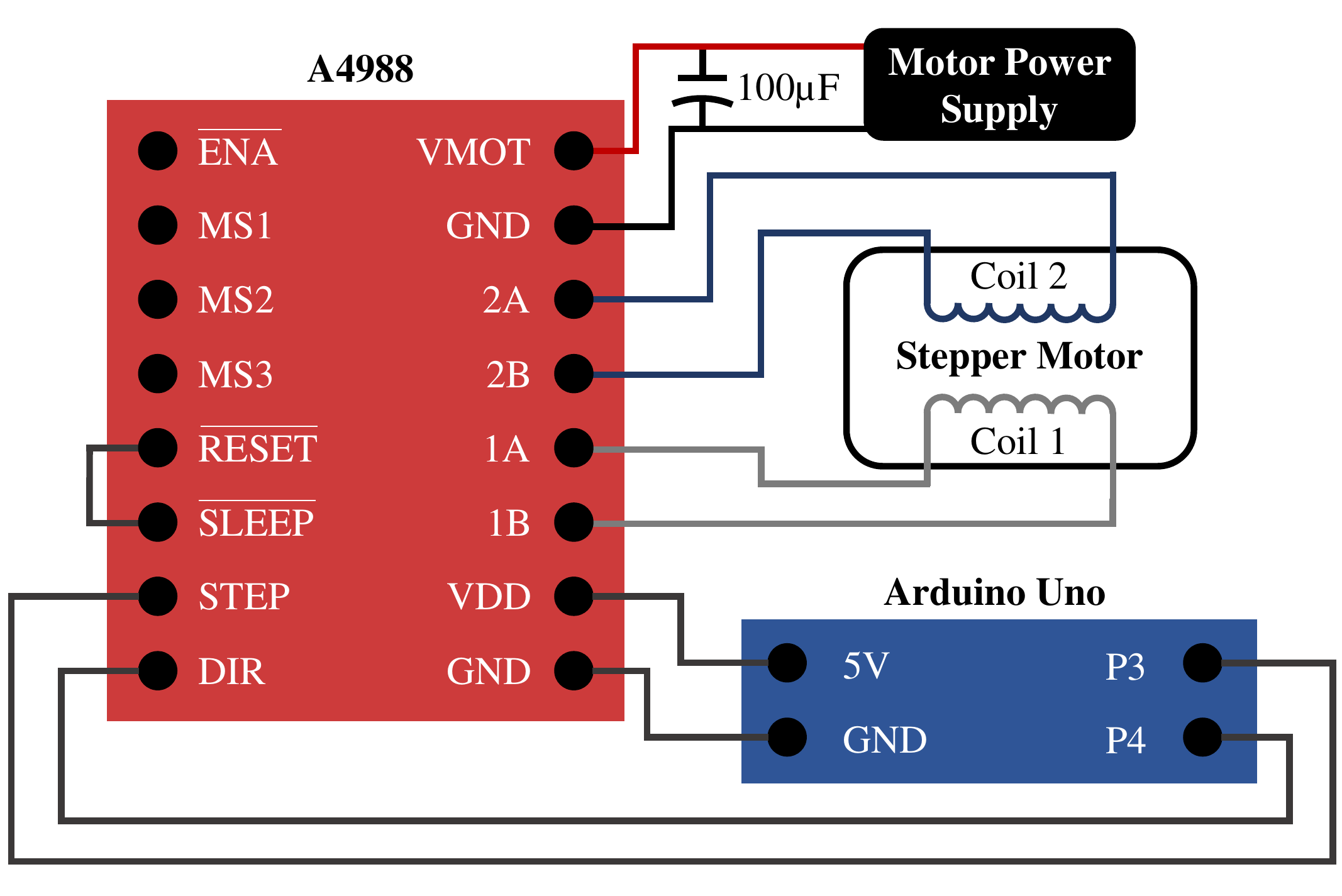}
\caption{Circuit diagram for stepper motor control using an Arduino Uno micro-controller and a motor driver module A4988.}
\label{fig:MotorControl}
\end{figure}

To operate the NEMA 17 stepper motor with \SI{0.22}{Nm} torque, three components are needed: a DC motor power supply, an Arduino Uno micro-controller and a motor driver module A4988 (Fig. \ref{fig:MotorControl}). In the first step, we connected the VMOT and the Ground pins of the driver module to a DC power supply, providing an voltage output of \SI{10}{\volt} to run the motor. To protect the A4988 driver board from voltage spikes, we shunted a decoupling capacitor (capacity $C=\SI{100}{\micro\farad}$) between the two power wires. The following four pins (2B, 2A, 1B, 1A) are used to connect the stepper motor with the A4988 driver card. For that purpose, we wired the pin couple 2A and 2B to coil 1 and couple 1A and 1B to coil 2 of the motor, respectively. The next two pins (VDD and Ground) are connected to the \SI{5}{\volt} and GND pin on the Arduino Uno card, being the power supply for the A4988 driver card. The step pin (STEP) and direction pin (DIR), controlling the step size and rotation direction of the motor, respectively, are wired to Arduino card pins 3 and 4. These two pins are gated by the Arduino algorithm provided in section \ref{sec:Appendix}\ref{subsec:ArduinoAlgorithm}. In the present setting, the stepper motor operates in full step mode. Eventually, we short-circuit the reset and sleep pin to avoid the stepper motor entering sleep mode. With the Arduino algorithm provided in section \ref{sec:Appendix}\ref{subsec:ArduinoAlgorithm} the stepper motor operates continuously at its maximum rotation speed of \SI{468}{\rpm}, corresponding to a rotation frequency of \SI{7.8}{\hertz}. With an intrinsic motor resonance frequency of \SIrange{0.5}{1.5}{\hertz}, the stepper motor runs smooth and stable at a rotation frequency of \SI{7.8}{\hertz}. 
\subsection{Sample preparation for validation experiments and microscope details}
\label{subsec:SamplePrep}
To carry out validation experiments, we prepare a solution of polystyrene (PS) mono size particles of \SI{1.040\pm 0.022}{\micro\meter} (mean $\pm$ SD), Lot No. 15879, Duke Scientific Corp., \SI{4}{\%} w/v) in deionized water or phosphate buffered saline (PBS, pH 7.4). Measurements are performed in a simple sandwich cell, consisting of two cover slips (lower cover slip: no. 1, Knittel Glass, \SI{60x24}{\milli\meter}; upper cover slip: no. 1, Knittel Glass, \SI{20x20}{\milli\meter}) separated by double sticky tape (Scotch, product no. 34-8509-3289-7, 3M). The cell volume is $\approx\SI{10}{\micro\liter}$. We inject the diluted bead solution with a micro-pipette and seal the chamber by adding vacuum grease (Dow Corning) on both openings to prevent evaporation, which can result in a biased movement of the particles. To investigate motor-induced vibrations (section \ref{sec:ValidationExp}\ref{subsec:Vibrations}) we incubate the sandwich cell overnight so PBS dissolved particles settle down and immobilize on the bottom cover slip. In contrast, measurements to measure the Brownian motion (section \ref{sec:ValidationExp}\ref{subsec:BrownianMotion}) are performed immediately after injecting particles dissolved in deionized water into the sample cell. 

Measurements are done using an Olympus IX70 inverted microscope, normally used for optical tweezers experiments, equipped with either an oil-immersion objective (Olympus UPlanFl $100\text{x}/1.30$ Oil, $\infty/0.17$) or an free space objective (Olympus SLMPLN $50\text{x}/0.35$, $\infty/0$) \cite{Fallman2004}. However, the 50x SLMPLN objective is only used to determine the highest camera frame rate at which speckle pattern reappear in the image background (section \ref{sec:ValidationExp}\ref{subsec:ImageQuality}). The prepared sample cell is mounted onto a piezo stage, which can be positioned in three dimensions over a range of \SI{100}{\micro\meter} with nanometer accuracy using piezo actuators (P-561.3CD, Physik Instrumente). The sample is illuminated from above with parallel light produced from the optical fiber collimator (Appendix \ref{sec:Appendix}\ref{subsec:OpticalComponents}), and particles are imaged using a high-speed camera (MotionBLITZ EoSens Cube 7, Mikrotron) operating at \SI{150}{\fps} (section \ref{sec:ValidationExp}\ref{subsec:Vibrations}) and \SI{2000}{\fps} (section \ref{sec:ValidationExp}\ref{subsec:BrownianMotion}). We acquire images using the MotionBLITZDirector2 software and the resolution of the microscopy system is \SI{80\pm2}{\nano\meter\per\pixel} (mean $\pm$ SD). The whole setup is built in a temperature controlled room at \SI{296\pm1}{\kelvin} to ensure long-term stability and to reduce thermal drift effects \cite{Andersson2011}. For image analysis, we use the software UmUTracker \cite{Zhang2017}.\\
\section{Demonstration experiments}
\label{sec:ValidationExp}
After aligning all optical components and controlling the rotation speed of the stepper motor with an Arduino card, we characterize the performance of our setup. First, we show that the rotating ground glass diffuser reduces speckle pattern providing high quality microscopic images (section \ref{sec:ValidationExp}\ref{subsec:ImageQuality}). Second, we discuss experimental settings how to collect the scattered light from the RGG to maximize the optical fiber intensity output (section \ref{sec:ValidationExp}\ref{subsec:OutputIntensity}). Third, we scrutinize if the rotating motor induces vibrations into the setup that might reduce the accuracy in vibration-sensitive measurements (section \ref{sec:ValidationExp}\ref{subsec:Vibrations}). Eventually to feature the high laser light throughput, we record the Brownian motion of a freely diffusing particle using a 100x oil immersion objective at a camera frame rate of \SI{2000}{\hertz} and determine its diffusion constant (section \ref{sec:ValidationExp}\ref{subsec:BrownianMotion}).
\subsection{Image quality}
\label{subsec:ImageQuality}
\begin{figure}[htbp]
\centering
\includegraphics[width=\linewidth]{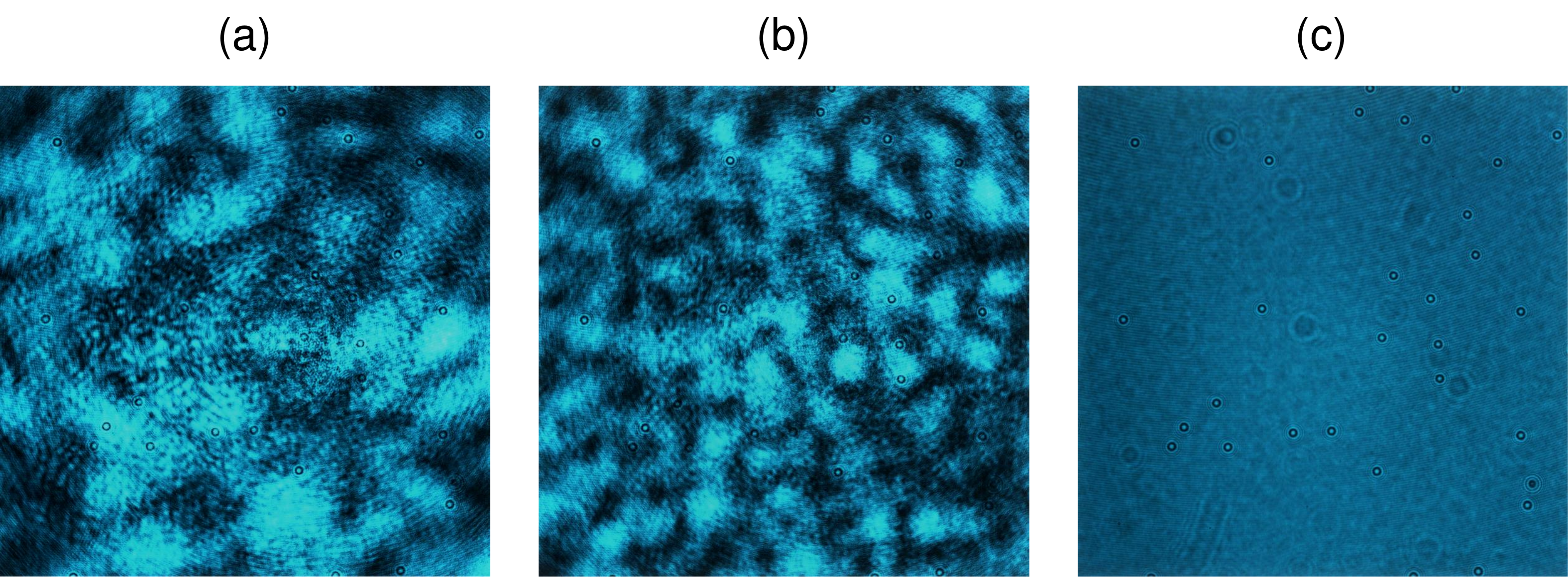}
\caption{Microscopic images acquired under different experimental conditions. (a) Speckle pattern produced by laser light without rotating ground glass diffuser. (b) Speckle pattern produced by laser light which is scattered by a stationary ground glass diffuser. (c) Image with rotating ground glass diffuser. Since the RGG reduce the spatial coherence of the laser light, no speckle pattern is visible.}
\label{fig:ImageQuality}
\end{figure}

Reducing the speckle pattern using a RGG depends strongly on experimental settings such as illuminated area on the ground glass diffuser, its grid size on the polished side, its rotation speed but also on the camera shutter time used to acquire images. If the laser beam is focused into the fiber entrance without ground glass diffuser, a characteristic speckle pattern can be observed (Fig. \ref{fig:ImageQuality} a). If a stationary ground glass diffuser is introduced into the beam path the speckle field pattern still appears, but its properties depend now on the grid size of the ground glass diffuser (Fig. \ref{fig:ImageQuality} b). By rotating the ground glass diffuser with a certain frequency, the speckle pattern disappears and an image without coherent noise and parasitic reflections can be recorded (Fig. \ref{fig:ImageQuality} c). Detailed descriptions about how the mentioned experimental parameters affect the coherence length of the laser and therefore the speckle pattern can be found elsewhere \cite{Asakura1970,Dubois2004,Estes1971}. 

In this context, we will now summarize and discuss experimental settings used for the presented setup to achieve efficient reduction of speckle pattern. We obtain the lowest speckle noise if the illumination area on the ground glass diffuser is maximized but small enough to ensure parallel light output from the collection objective. In the presented setup the illuminated area on the RGG has a diameter $d_{\text{ill. area}}\approx\SI{60}{\micro\meter}$ (Table \ref{tab:ComparisonObjective}). We will further discuss this point in section \ref{sec:ValidationExp}\ref{subsec:OutputIntensity}. 

By considering the surface properties of ground glass diffusers, the most efficient way to reduce speckle patterns is to use the smallest grid size available \cite{Asakura1970}. We choose a grid size of 1500. In contrast to the illuminated area on the ground glass diffuser and its grid size, the rotation speed plays only a minor role in reducing the spatial coherence after reaching the threshold rotation speed \cite{Asakura1970}. This value is in our setup \SI{400}{\rpm} for camera shutter times down to \SI{100}{\micro\second}. Therefore, we run the stepper motor at its maximum speed of \SI{468}{\rpm}. Since the laser beam hits the RGG at \SI{2.24}{\milli\meter} above the rotation center this corresponds to an tangential velocity of \SI{110}{\milli\meter\per\second}. With this experimental setting, speckle-free images can be acquired up to \SI{10000}{\hertz} (shutter time: \SI{100}{\micro\second}). Above this frame rate speckle pattern reappear in the background of the image. To image at higher frame rates or shorter shutter times, the stepper motor has to operate at higher rotation frequency to ensure that speckle pattern are effectively reduced. However, as mentioned in section \ref{sec:MAM}\ref{sec:Motor} the motor used in this setup has its maximum at \SI{468}{\rpm}, necessitating the usage of a stepper motor with higher torque.
\subsection{Choice of collection objective and output intensity}
\label{subsec:OutputIntensity}
Besides removing speckle pattern effectively, a high fiber intensity output to guaranty enough light for measurements at high camera frame rates is desirable. As the RGG scatters incoming light significantly, the closer to the scattering surface the light is collected, the higher the collection efficiency is. Therefore, we use an ordinary microscope objective instead of normal lenses since small distances between RGG and objective can be realized. The only requirement for the used collection objective (CO, Fig. \ref{fig:Setup} b) is a large front aperture to ensure effective light collection. For that reason, we choose a standard objective (Olympus Plan N 10x). To judge its collection efficiency, we determine the beam power to be \SI{10.8}{\milli\watt} in front of the RGG and measure the fiber output power using a power meter, revealing a fiber output power of \SI{5.2}{\milli\watt} (Table \ref{tab:ComparisonObjective}). This efficient light throughput of \SI{48}{\%} enables us to perform imaging up to \SI{10000}{\hertz} at only tens of Milliwatts laser power, preventing the setup and sample from laser-induced damaged. 

To achieve optimal light throughput we ensure that the output beam from the collection objective is parallel since this beam will be guided into the optical fiber by the focusing objective. However, objectives produce only collimated light under certain conditions. We will clarify this in the following. Consider a laser beam with a Gaussian intensity distribution entering the back aperture of an objective. If the beam diameter equals the objectives entrance pupil diameter, the light will be focused at the working distance of the objective. The spot size can then be approximated by the diffraction limit: $d_{\text{Spot}}\approx 1.22 \lambda/\text{N.A.}$ which is $\approx\SI{3}{\micro\meter}$ for the 10x collection objective (Table \ref{tab:ComparisonObjective}). Vice versa, an objective produces a collimated beam if the diameter of the imaged light source is less than or equal to $d_{\text{Spot}}$. In other words, if the diameter of the illuminated area ($d_{\text{ill. area}}$) on the RGG is bigger than $d_{\text{Spot}}$, the output beam from the collection objective will be divergent. The divergence becomes more pronounced the larger the ratio between $d_{\text{ill. area}}$ and $d_{\text{Spot}}$ differs from unity. The same condition applies if a lens is used for light collection (see next paragraph and Table \ref{tab:ComparisonObjective}). To meet the discussed requirement for collimated beam output, we introduced lens L1 into the light path to reduce the diameter of the illuminated are on the RGG from \SI{1300}{\micro\meter} to \SI{60}{\micro\meter}. In this case, the collection objective produces slightly divergent light. To compensate the divergence, we place the collection objective marginally closer to the RGG than its working distance. Despite focusing the laser beam on the RGG, the illumination area is still big enough to realize effective speckle averaging for camera frame rates up to \SI{10 000}{\hertz}.

To highlight the high light collection efficiency by using only one objective, we compare this scenario to a single lens collector as reported in literature \cite{Dubois2004,Fougeres1994,Neil2000,Scarcelli2004,Wang2006}. For that purpose, we place a \SI{60}{\milli\meter} lens (LA1134-A; Thorlabs) behind the ground glass diffuser, possessing a similar numerical aperture (N.A.) as the used objective (Table \ref{tab:ComparisonObjective}). Again, the laser intensity in front of the RGG is set to \SI{10.8}{\milli\watt}. The lens output is a slightly divergent beam (see discussion above) with a diameter of \SI{25.4}{\milli\meter} and an intensity of \SI{4.4}{\milli\watt}. Guiding this beam into the fiber produces an output of only \SI{0.2}{\milli\watt} as the majority of the light is truncated by the back aperture of the focusing objective. Therefore, we decrease the beam size using a negative beam magnification (LA1509-A, LA1134-A; Thorlabs), revealing an fiber output of \SI{3.1}{\milli\watt}, which is $\approx\SI{40}{\%}$ lower than the achieved intensity using an objective for light collection. Furthermore, using the same objective for light collection and light focusing into the fiber has the advantage, that no further optics are needed to manipulate the beam diameter since the collimated output beam has the same diameter as the back aperture of the focusing objective. 
\begin{table}[htbp]
\centering
\caption{\bf Comparison of fiber intensity output for two light collecting options: objective and single lens (WD = working distance, N.A. = numerical aperture, $d_{\text{Spot}}$ = theoretical focal spot diameter, $d_{\text{ill. area}}$ = diameter of illuminated area on RGG, $P_{\text{x}}$ = laser intensity).}
\begin{tabular}{ccc}
\hline
Objective						& Olympus Plan N 10x 				& single lens\\
\hline
WD 								& \SI{10.6}{\milli\meter}			& \SI{60}{\milli\meter} \\
N.A.							& \SI{0.25}{}						& \SI{0.21}{}\\
$d_{\text{Spot}}$				& \SI{3.0}{\micro\meter}				& \SI{35}{\micro\meter}\\
$d_{\text{ill. area}}$			& \SI{60}{\micro\meter}				& \SI{60}{\micro\meter}\\
$P_{\text{before RGG}}$ 		& \SI{10.8}{\milli\watt} 			& \SI{10.8}{\milli\watt} \\
$P_{\text{fiber output}}$		& \SI{5.2}{\milli\watt}				& \SI{3.1}{\milli\watt}\\
coupling eff.					& $\approx\SI{48}{\%}$				& $\approx\SI{29}{\%}$\\
\hline
\end{tabular}
  \label{tab:ComparisonObjective}
\end{table}
\subsection{Vibrations caused by the RGG do not impair image quality}
\label{subsec:Vibrations}
\begin{figure}[htbp]
\centering
\includegraphics[width=\linewidth]{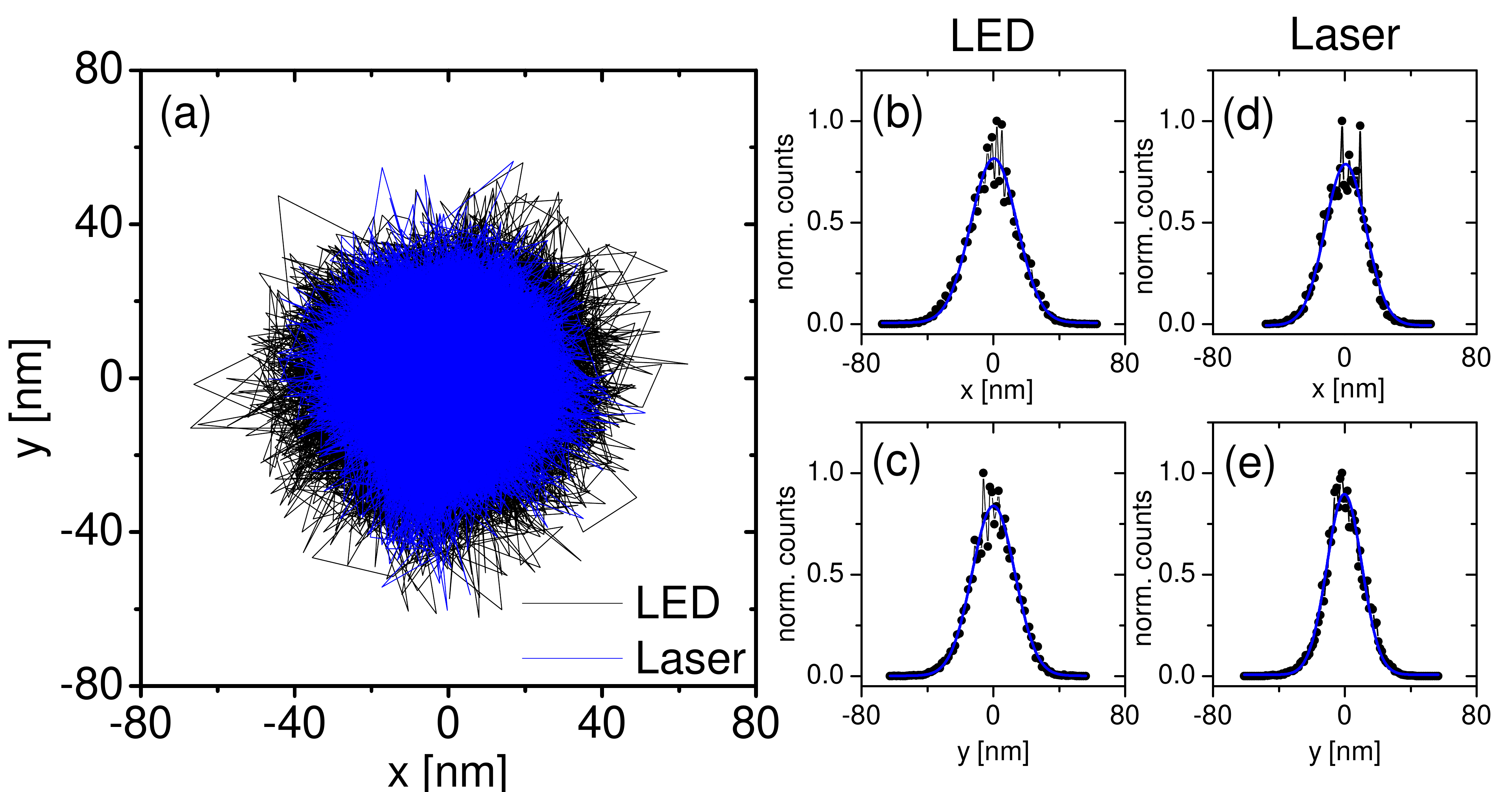}
\caption{Influence of vibrations caused by the RGG when determining the particle center position. (a) Comparison of particle center positions estimated under LED (without RGG, black data) and laser (with RGG, blue data) illumination acquired at a frame rate of \SI{150}{\hertz} (camera shutter time \SI{6.7}{\milli\second}). (b-c) Quantitative analysis of particle movement in the \textit{x} and \textit{y} direction under LED illumination (black data points). A Gaussian function (blue line) is fitted to the data, revealing a full width at half maximum of \SI{33}{\nano\meter} (\textit{x} direction, $R^2 = 0.97$) and \SI{30}{\nano\meter} (\textit{y} direction, $R^2 = 0.98$), respectively. (d-e) Position analysis of the same particle as in (b-c) but with the RGG and laser illumination (black data points). The Gaussian fit (blue line) produces a full width at half maximum of \SI{29}{\nano\meter} (\textit{x} direction, $R^2 = 0.96$) and \SI{25}{\nano\meter} (\textit{y} direction, $R^2 = 0.98$). }
\label{fig:Vibrations}
\end{figure}

In case the gear and the ground glass diffuser are not absolutely centered to each other (section \ref{sec:Appendix}\ref{subsec:InstallationDetails}), the RGG can induces vibrations that might perturb sensitive measurements. This problem becomes pronounced if the stepper motor is placed on the same optical table as the measurement microscope. To estimate the impact of vibrations on measurements when the stepper motor is running at its maximum rotation speed, we immobilize \SI{1}{\micro\meter} PS particles on the bottom cover slip of the sandwich chamber (section \ref{sec:MAM}\ref{subsec:SamplePrep}) and image their position using laser (with RGG) and LED (without RGG) illumination at \SI{150}{\hertz} (Fig. \ref{fig:Vibrations} a). To compare the position width using laser (blue data points) and LED illumination (black data points), we analyze their profiles along the \textit{x} and \textit{y} direction using a Gaussian fit (Fig. \ref{fig:Vibrations} b-e). As a result we obtain a mean full width at half maximum (FWHM) of \SI{31.5\pm1.5}{\nano\meter} in \textit{x} and \SI{27\pm2}{\nano\meter} in \textit{y} direction, respectively. By considering a camera pixel size of \SI{80}{\nano\meter\per\pixel}, the bead position varies within one pixel and no difference in the particle center position using either laser or LED illumination can be found. Consequently, rotation induced vibrations at a rotation speed of \SI{468}{\rpm} can be neglected. 

Eventually, it should be mentioned that stepper motors possess a resonance frequency at which they produce heavy vibrations. For example, the motor used in this setup has its resonance frequency around \SIrange{0.5}{1.5}{\hertz}. At this rotation speed, vibrations will propagate to the sample and can be measured. Therefore, if the stepper motor is operated around its resonance frequency, we recommend to build the laser illumination setup with the RGG on a separate optical table, since longer distances can convincingly be bridged using a longer optical fiber.
\subsection{Brownian motion analysis}
\label{subsec:BrownianMotion}
\begin{figure}[htbp]
\centering
\includegraphics[width=\linewidth]{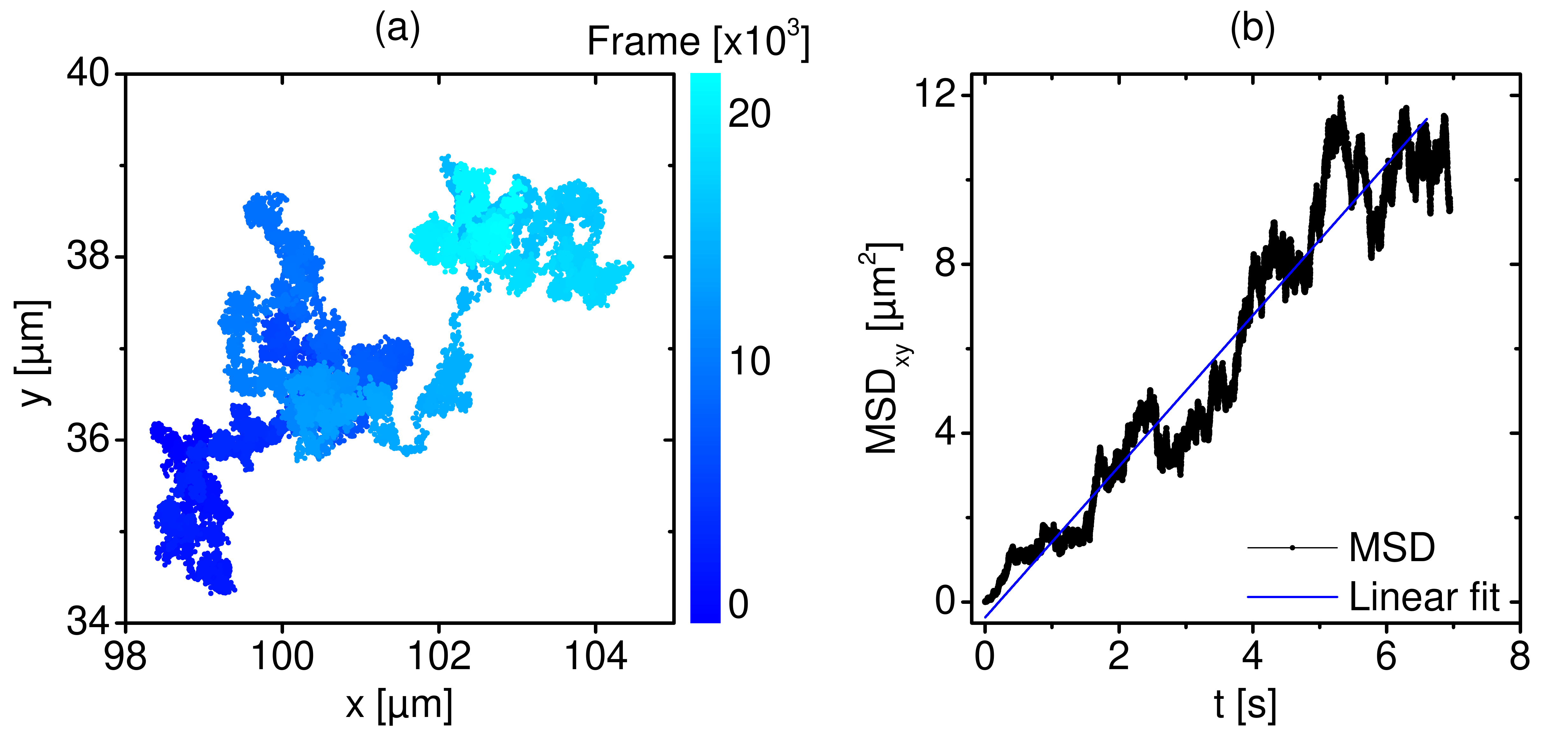}
\caption{(a) Typical 2D trajectory of a PS particle with a diameter of \SI{1}{\micro\meter} freely diffusing in deionized water. Data acquired at a frame rate of \SI{2000}{\hertz} resulting in 23000 data points. (b) 2D mean square displacement (black data points) versus time. To calculate the MSD four independent particle trajectories are used. By fitting the MSD with a linear function (blue line, $R^2 = 0.94$) the diffusion constant is assessed to $D = \SI{0.45\pm0.05}{\micro\meter\squared\per\second}$.}
\label{fig:BM_MSD}
\end{figure}

High speed imaging requires short shutter times, which in turn demands an adequate amount of light reaching the sample under study. To highlight the potential of our speckle-free laser illumination setup, we track freely diffusing \SI{1}{\micro\meter} PS particles in deionized water in two dimensions (Fig. \ref{fig:BM_MSD} a). To achieve optimal spatial resolution, we use a 100x oil immersion objective and acquire images at a camera frame rate of \SI{2000}{\hertz} (camera shutter time \SI{0.5}{\milli\second}). The laser light output at the fiber collimator is set to \SI{22}{\milli\watt}. From the trajectories of four independent diffusing particles, we calculate the 2D mean square displacement (MSD, Fig. \ref{fig:BM_MSD} b, black data) and fit the data with a linear function (blue line). By considering the relation between MSD and diffusion constant $D$: $\text{MSD}(\Delta t)=4D\Delta t$, the slope of the linear fit equals $4D$ revealing a diffusion constant of $D=\SI{0.45\pm0.05}{\micro\meter\squared\per\second}$. This value can be theoretically validated by using the Einstein-Stokes relation \cite{Einstein1905}: $D=k_{\text{B}}T\cdot\left(3\pi\eta d\right)^{-1}$, with $k_{\text{B}}T$ being the thermal energy, $\eta$ the viscosity of water and $d$ the particle diameter. With a temperature $T=\SI{296}{\kelvin}$, a viscosity $\eta=\SI{0.932e-3}{\pascal\second}$ and a particle diameter $d=\SI{1.04}{\micro\meter}$, we calculate the theoretical diffusion constant to $D_{\text{theo}}=\SI{0.45}{\micro\meter\squared\per\second}$, showing agreement with our experimental value. 
\section{Conclusion}
\label{sec:Conculsion}
We presented a step-by-step guide to build a speckle-free laser illumination setup using a rotating glass diffuser mounted on a stepper motor, that is easy to realize, compact and robust. Furthermore, by using only one objective to collect the scattered light behind the ground glass diffuser, we improved the light throughput by $\approx \SI{40}{\%}$ compared to using single lenses collector. Costs for the stepper motor, its controlling unit and the objective are below 500 USD being significantly cheaper than commercial solutions. We provided detailed experimental information for setup construction, enabling the user to record speckle-free images up to \SI{10000}{\hertz}. We characterized our setup in terms of image quality, light throughput and motor-induced vibrations. To highlight a light throughput of $\approx\SI{48}{\%}$, we recorded the Brownian motion of micro-particles at a frame rate of \SI{2000}{\hertz} using a 100x oil immersion objective and determined the diffusion constant. By providing microscopic images with highest quality at high camera frame rates and moderate laser intensities, our setup can be used for example to investigate surface-attached bacteria under physiological flow conditions using digital holography without applying image post-processing such as background subtraction or filtering.
\section{Funding Information}
T.S. acknowledges financial support from the German Research Foundation (DFG) via a postdoctoral fellowship. This work was supported by the Swedish Research Council (2013-5379) and from the Kempe foundation to M.A.
\section{Appendix: Component lists, Installation details and Arduino algorithm}
\label{sec:Appendix}
\subsection{Optical Components}
\label{subsec:OpticalComponents}
\begin{enumerate}
\item Cobolt Calypso\texttrademark 50 laser ($\lambda= \SI{491}{\nano\meter}$, max. output power $P=\SI{45}{\milli\watt}$), Cobolt AB, Solona, Sweden -- Fig. \ref{fig:Setup}: Laser
\item Optical beam shutter, Thorlabs (SH05) -- Fig \ref{fig:Setup}: Shutter
\item Kinematic mirror mount for $\oslash1^{\prime\prime}$ optics, Thorlabs (KM100) -- Fig. \ref{fig:Setup}: M1, M2, L1 \& Fig. \ref{fig:AppendixRGG}: Mount
\item 30 mm cage plate for $\oslash1^{\prime\prime}$ optics, Thorlabs (CP06) 
\item $\oslash1^{\prime\prime}$ broad band dielectric mirror, \SIrange{400}{750}{\nano\meter}, Thorlabs (BB1-E02) -- Fig. \ref{fig:Setup}
\item $\oslash1^{\prime\prime}$ N-BK7 plano-convex lens (AR coating: \SIrange{350}{700}{\nano\meter}), $f=\SI{100}{\milli\meter}$, Thorlabs (LA1509-A) -- Fig. \ref{fig:Setup}: L1
\item $\oslash2^{\prime\prime}$ unmounted N-BK7 ground glass diffuser, 1500 grit, Thorlabs (DG20-1500) -- Fig. \ref{fig:Setup} \& Fig. \ref{fig:AppendixRGG}: RGG
\item 10x Olympus plan achromat objective, 0.25 NA, \SI{10.6}{\milli\meter} WD, Thorlabs (RMS10X), -- Fig. \ref{fig:Setup} \& Fig. \ref{fig:AppendixRGG}: CO and FO
\item Fiber launch with FC-connectorized fiber holder, Thorlabs (MBT613D)
\item $\oslash\SI{400}{\micro\meter}$, 0.39 NA, FC/PC to SMA multimode hybrid fibre optic patch cable, Thorlabs (M76L02)  -- Fig \ref{fig:Setup} (OF)
\item Aspheric FiberPort, SMA, $f=\SI{7.5}{\milli\meter}$, \SIrange{350}{700}{\nano\meter}, $\oslash\SI{1.23}{\milli\meter}$ waist, Thorlabs (PAF-SMA-7-A)
\end{enumerate}
\subsection{Mechanical Components}
\label{subsec:MechnaicalComponents}
\begin{itemize}
\item Triple divide \textit{xyz} translation stage, \SI{28}{\milli\meter} travel, Newport (9064-XYZ-M) -- Fig. \ref{fig:Setup} \& Fig. \ref{fig:AppendixRGG}: TS
\item NEMA 17 high torque stepper motor, 42H33H-1334A, \SI{0.22}{\newton\meter} torque, Amazon -- Fig. \ref{fig:Setup} \& Fig. \ref{fig:AppendixRGG}: SM
\item Aluminium V-belt pulley Reely bore diameter: \SI{5}{\milli\meter}, outer diameter: \SI{20}{\milli\meter}, Conrad (238325-62) -- Fig. \ref{fig:Setup} \& Fig. \ref{fig:AppendixRGG}: Gear
\item Switchable, heavy-duty magnetic base, Thorlabs (MB175/M) -- Fig. \ref{fig:Setup} \& Fig. \ref{fig:AppendixRGG}: Base
\item 3D printed motor holder, CAD file and mechanical drawing can be downloaded from Ref. \cite{Stangner2017} -- Fig. \ref{fig:Setup} \& Fig. \ref{fig:AppendixRGG}: MH 
\item 3D printed holder for $\oslash1/2^{\prime\prime}$ stainless steel optical posts, CAD file and mechanical drawing can be downloaded from Ref. \cite{Stangner2017} -- Fig. \ref{fig:Setup} \& Fig. \ref{fig:AppendixRGG}: PH
\item 3D printed holder for RMS threated objectives, CAD file and mechanical drawing can be downloaded from Ref. \cite{Stangner2017} -- Fig. \ref{fig:Setup} \& Fig. \ref{fig:AppendixRGG}: OR 
\end{itemize}
\subsection{Electrical Components}
\label{subsec:ElectricalComponents}
\begin{itemize}
\item GW Instek GPS-30300 DC power supply, Gw Instek
\item Arduino Uno R3 micro-controller, Amazon -- Fig. \ref{fig:MotorControl}
\item Stepstick stepper motor driver module A4988, Amazon -- Fig. \ref{fig:MotorControl}
\item Breadboard, K\&H MFG. Co., LTD (RH-21B) -- Fig. \ref{fig:MotorControl}
\item Panasonic \SI{100}{\micro\farad} radial electrolyte capacitor, Conrad (421945-62) -- Fig. \ref{fig:MotorControl} 
\end{itemize}
\subsection{Installation details for stepper motor with rotating ground diffuser and collection objective}
\label{subsec:InstallationDetails}
\begin{figure}[htbp]
\centering
\includegraphics[width=\linewidth]{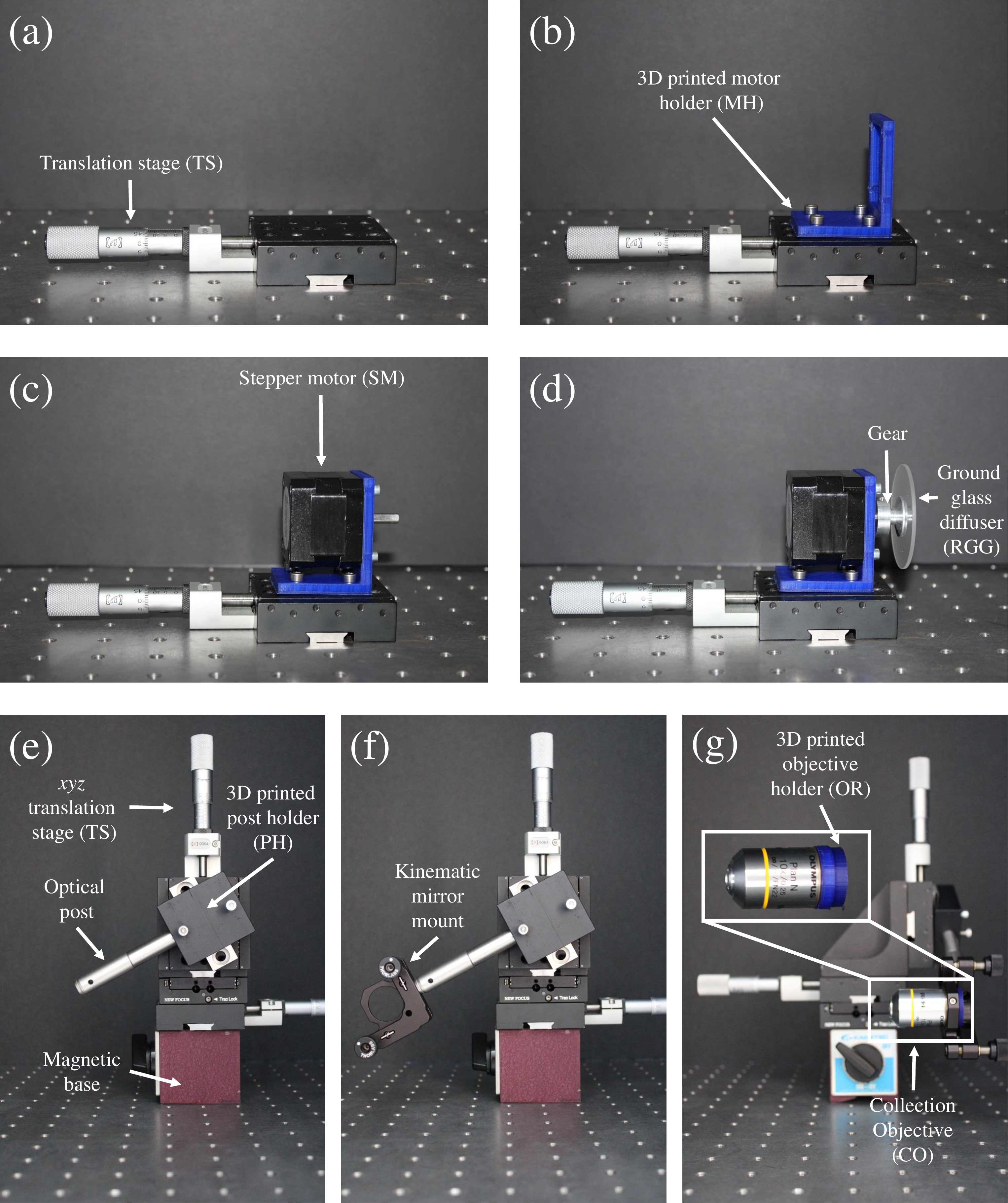}
\caption{Step-by-step manual to assemble the stepper motor (SM) with the ground glass diffuser (RGG) and the collection objective (CO) with its holder (OR).}
\label{fig:AppendixRGG}
\end{figure}

In this section, we provide step-by-step information how to install the stepper motor (SM) on the translation stage and how to glue the ground glass diffuser (RGG) to a gear, which will be mounted on the D-cut motor shaft. Furthermore, we explain how the collection objective (CO) is installed on the \textit{xyz}-translations stage (Fig. \ref{fig:AppendixRGG}).

We start with the assembly of stepper motor and ground glass diffuser. In the first step, a single translation stage is mounted on an optical table (Fig. \ref{fig:AppendixRGG} a). Second, the lower surface of the 3D printed motor holder is screwed to the stage using four screws (Fig. \ref{fig:AppendixRGG} b). The CAD file and the mechanical drawing file of the holder can be downloaded from Ref. \cite{Stangner2017}. Third, the stepper motor (SM) is mounted to the front side of the holder (Fig. \ref{fig:AppendixRGG} c). Before tighten the gear on the D-cut motor shaft with a set screw (Fig. \ref{fig:AppendixRGG} d), we glue the ground glass diffuser with its smooth surface to the gear. For this we use a cyanoacrylate-based adhesive. The gluing step is crucial and must be carried out with highest precision to ensure that gear and ground glass diffuser are centered. If the centers are misaligned, the motor will be imbalanced and vibrations at high rotation speeds occur.

To control the position and orientation of the collection objective (CO) using a \textit{xyz}-translation stage, the latter is pre-assembled on a magnetic base containing a switchable magnet. Afterwards, the 3D printed holder for $\oslash1/2^{\prime\prime}$ stainless steel optical posts and the post are mounted on the \textit{z}-translation stage (Fig. \ref{fig:AppendixRGG} e). In the next step, we mount a kinematic mirror mount (Mount) on the tip of the post (Fig. \ref{fig:AppendixRGG} f) to control the tilt of the collection objective. Thereafter, we pre-assemble the collection objective (CO) and the 3D printed holder for RMS threated objectives (OR) (Fig. \ref{fig:AppendixRGG} g, Inset). In the final step, we insert the objective holder into the kinematic mirror mount and tighten it with a set screw (Fig. \ref{fig:AppendixRGG} g).
\subsection{Arduino Algorithm}
\label{subsec:ArduinoAlgorithm}
Here, we provide the Arduino algorithm \ref{alg:arduino} to control the rotation speed of the stepper motor (NEMA 17, section \ref{sec:MAM}\ref{sec:Motor}), written in Arduino Software 1.8.1. With the provided settings, the motor runs at \SI{468}{\rpm}. However, to achieve this maximum rotation speed, the stepper motor needs to accelerate slowly by starting at a low speed. This is caused by the motor inertia and the mass of the ground glass diffuser.
The first 3 lines of the code define motor-dependent input values, such as initial time for one revolution ("\textit{const int InitialTimeStep = 700}"), acceleration time step ("\textit{const int AccelerationTimeStep = 2;}") and time per revolution after the final acceleration step ("\textit{const int FinalTimeStep= 128;}" in [\SI{}{\milli\second}]). The latter sets the maximum rotation speed. Afterwards, the step ("\textit{stepPin}") and direction ("\textit{dirPin}") pin on the Arduino board are specified and set as output. The loop starting in line 13 controls the acceleration process based on the three predefined input values mentioned above. Once the acceleration has reached the "\textit{FinalTimeStep}", the motor will run with this value for \SI{1000}{\minute} (line 24). The algorithm code can be downloaded from Ref. \cite{Stangner2017}.
\begin{algorithm}[htbp]
\caption{Arduino algorithm to control the rotation speed of a stepper motor.}
\label{alg:arduino}
\begin{algorithmic}[1]
\State{const int InitialTimeStep =700;}
\State{const int AcceleratingTimeStep= 2;}
\State{const int FinalTimeStep= 128;}\\
\State{const int stepPin = 3;}				\Comment{defines pin numbers}
\State{const int dirPin = 4;}\\
\State {\textbf{void} setup() \{ }			\Comment{sets the two pre-defined pins as output}
	\State{pinMode(stepPin, OUTPUT);}
	\State{pinMode(dirPin, OUTPUT); }
    \State{\}}\\
\State{\textbf{void} loop()} \{				\Comment{Acceleration part}
	\State{digitalWrite(dirPin, HIGH);}
	\State{\textbf{for} (int i = InitialTimeStep; i > FinalTimeStep; i = i-AcceleratingTimeStep)\{)}
    \State{\textbf{for} (int x = 0; x < 200; x++) \{}
	  	\State{digitalWrite(stepPin, HIGH)};
		\State{delayMicroseconds(i);}
		\State{digitalWrite(stepPin, LOW);}
		\State{delayMicroseconds(i);}
		\State{\}}
		\State{\}}\\
\State{\textbf{for} (int i = InitialTimeStep; i < 3600000; i = i++) \{} \Comment{Defines run time of script}
\State{\textbf{for} (int x = 0; x < 200; x++) \{}
		\State{digitalWrite(stepPin, HIGH)};
		\State{delayMicroseconds(FinalTimeStep);}
		\State{digitalWrite(stepPin, LOW);}
		\State{delayMicroseconds(FinalTimeStep);}
        \State{\}}
	\State{\}}
\State{\}}
\end{algorithmic}
\end{algorithm}
%

\begin{thebibliography}{10}
	\newcommand{\enquote}[1]{``#1''}
	
	\bibitem{DaintyJ.C.1975}
	{Dainty J. C.}, \enquote{{Laser speckle and related phenomena},} Topics in
	Applied Physics \textbf{9}, 298 (1975).
	
	\bibitem{Goodman1976}
	J.~W. Goodman, \enquote{{Some fundamental properties of speckle},} Journal of
	the Optical Society of America \textbf{66}, 1145--1150 (1976).
	
	\bibitem{Goodman2005}
	J.~W. Goodman, \emph{{Speckle phenomena in optics: theory and applications}}
	(2005).
	
	\bibitem{Gratton2006}
	E.~Gratton and M.~J. VandeVen, \enquote{{Laser Sources for Confocal
			Microscopy},} Handbook of Biological Confocal Microscopy pp. 80--125 (2006).
	
	\bibitem{alphonse1989super}
	G.~A. Alphonse and D.~B. Gilbert, \enquote{{Super-luminescent diode},}  (1989).
	
	\bibitem{Goldberg1994}
	L.~Goldberg and D.~Mehuys, \enquote{{High power superluminescent diode
			source},} Electronic Letters \textbf{30}, 1682--1684 (1994).
	
	\bibitem{smith2011laser}
	D.~K. Smith and J.~A. Casey, \enquote{{Laser-driven light source},}  (2011).
	
	\bibitem{Redding2012}
	B.~Redding, M.~a. Choma, and H.~Cao, \enquote{{Speckle-free laser imaging using
			random laser illumination},} Nat. Photonics \textbf{6}, 355--359 (2012).
	
	\bibitem{Ellis1979}
	G.~W. Ellis, \enquote{{A fiber-optic phase-randomizer for microscope
			illumination by laser},} Journal of Cell Biology \textbf{83}, 303a (1979).
	
	\bibitem{Davenport1992}
	C.~M. Davenport and A.~F. Gmitro, \enquote{{Angioscopic fluorescence imaging
			system},}  (1992), vol. 1649, pp. 192--202.
	
	\bibitem{Bains1993}
	S.~Bains, \enquote{{Holographic optics: for when less is more},} Laser focus
	world \textbf{29}, 151--156 (1993).
	
	\bibitem{Asakura1970}
	T.~Asakura, \enquote{{Spatial coherence of laser light passed through rotating
			ground glass},} Opto-electronics \textbf{2}, 115--123 (1970).
	
	\bibitem{Estes1971}
	L.~E. Estes, L.~M. Narducci, and R.~A. Tuft, \enquote{{Scattering of Light from
			a Rotating Ground Glass},} Journal of the Optical Society of America
	\textbf{61}, 1301--1306 (1971).
	
	\bibitem{Kumar2010}
	A.~Kumar, J.~Banerji, and R.~P. Singh, \enquote{{Intensity correlation
			properties of high-order optical vortices passing through a rotating
			ground-glass plate.}} Optics letters \textbf{35}, 3841--3 (2010).
	
	\bibitem{Zhai2005}
	Y.~H. Zhai, X.~H. Chen, D.~Zhang, and L.~A. Wu, \enquote{{Two-photon
			interference with true thermal light},} Physical Review A - Atomic,
	Molecular, and Optical Physics \textbf{72} (2005).
	
	\bibitem{Shapiro2008}
	J.~H. Shapiro, \enquote{{Computational ghost imaging},} Physical Review A -
	Atomic, Molecular, and Optical Physics \textbf{78}, 1--4 (2008).
	
	\bibitem{Ferri2005}
	F.~Ferri, D.~Magatti, A.~Gatti, M.~Bache, E.~Brambilla, and L.~A. Lugiato,
	\enquote{{High-resolution ghost image and ghost diffraction experiments with
			thermal light},} Physical Review Letters \textbf{94}, 2--5 (2005).
	
	\bibitem{Zhang2007}
	M.~Zhang, Q.~Wei, X.~Shen, Y.~Liu, H.~Liu, J.~Cheng, and S.~Han,
	\enquote{{Lensless Fourier-transform ghost imaging with classical incoherent
			light},} Physical Review A - Atomic, Molecular, and Optical Physics
	\textbf{75}, 1--4 (2007).
	
	\bibitem{Funatsu1995}
	T.~Funatsu, Y.~Harada, M.~Tokunaga, K.~Saito, and T.~Yanagida,
	\enquote{{Imaging of single fluorescent molecules and individual {\{}ATP{\}}
			turnovers by single myosin molecules in aqueous solution.}}  (1995).
	
	\bibitem{Dubois2004}
	F.~Dubois, M.-L. {Novella Requena}, C.~Minetti, O.~Monnom, and E.~Istasse,
	\enquote{{Partial spatial coherence effects in digital holographic microscopy
			with a laser source},} Appl. Opt. \textbf{43}, 1131--1139 (2004).
	
	\bibitem{Fougeres1994}
	A.~Fougeres and W.~Noh, \enquote{{Measurement of phase differences between two
			partially coherent coherent fields},} Physical Review A - Atomic, Molecular,
	and Optical Physics \textbf{49}, 530--535 (1994).
	
	\bibitem{Neil2000}
	M.~A.~A. Neil, A.~Squire, R.~Ju{\v{s}}kaitis, P.~I.~H. Bastiaens, and
	T.~Wilson, \enquote{{Wide-field optically sectioning fluorescence microscopy
			with laser illumination},} Journal of Microscopy \textbf{197}, 1--4 (2000).
	
	\bibitem{Scarcelli2004}
	G.~Scarcelli, A.~Valencia, and Y.~Shih, \enquote{{Experimental study of the
			momentum correlation of a pseudothermal field in the photon-counting
			regime},} Physical Review A - Atomic, Molecular, and Optical Physics
	\textbf{70}, 1--4 (2004).
	
	\bibitem{Wang2006}
	W.~Wang, Z.~Duan, S.~G. Hanson, Y.~Miyamoto, and M.~Takeda,
	\enquote{{Experimental study of coherence vortices: Local properties of phase
			singularities in a spatial coherence function},} Physical Review Letters
	\textbf{96}, 1--4 (2006).
	
	\bibitem{Trisnadi2002}
	J.~I. Trisnadi, \enquote{{Speckle contrast reduction in laser projection
			displays},} Proceedings of the Society of Photo-Optical Instrumentation
	Engineers (Spie) \textbf{4657}, 131--137 (2002).
	
	\bibitem{Fallman2004}
	E.~F{\"{a}}llman, S.~Schedin, J.~Jass, M.~Andersson, B.~E. Uhlin, and O.~Axner,
	\enquote{{Optical tweezers based force measurement system for quantitating
			binding interactions: system design and application for the study of
			bacterial adhesion},} Biosensors and Bioelectronics \textbf{19}, 1429--1437
	(2004).
	
	\bibitem{Andersson2011}
	M.~Andersson, F.~Czerwinski, and L.~B. Oddershede, \enquote{{Optimizing active
			and passive calibration of optical tweezers},} J. Opt. \textbf{13}, 044020
	(2011).
	
	\bibitem{Zhang2017}
	H.~Zhang, T.~Stangner, K.~Wiklund, A.~Rodriquez, and M.~Andersson,
	\enquote{{UmUTracker: A versatile MATLAB program for automated particle
			tracking of 2D light microscopy or 3D digital holography data},}   (2017).
	
	\bibitem{Einstein1905}
	A.~Einstein, \enquote{{{\"{U}}ber die von der molekularkinetischen Theorie der
			W{\"{a}}rme geforderte Bewegung von in ruhenden Fl{\"{u}}ssigkeiten
			suspendierten Teilchen},} Annalen der Physik \textbf{322}, 549--560 (1905).
	
\bibitem{Stangner2017}
T.~Stangner, \enquote{{Supplementary Information},}
\url{https://doi.org/10.6084/m9.figshare.4758451.v1} (2017). [Retrieved: 17
March 2017].

	
\end{thebibliography}

\end{document}